\documentclass{PoS}
\newcommand{\be}{\begin{eqnarray}}
\newcommand{\ee}{\end{eqnarray}}

\newcommand{\bfk}{{\bf k}_{\perp}}

\newcommand{\bfp}{{\bf p}_{\perp}}

\newcommand{\bfPhp}{{\bf P}_{h\perp}}
\newcommand{\bfPh}{{\bf P}_{h}}

\newcommand{\avpsq}{\langle{p^2}_{\perp}\rangle}
\newcommand{\avksq}{\langle{k^2}_{\perp}\rangle}
\newcommand{\avPhsq}{\langle{P^2}_{h\perp}\rangle}
\title{Proton structure in a light-front quark-diquark model: Collins asymmetry}

\ShortTitle{Collins asymmetry in LFQDM}

\author{\speaker{Tanmay Maji}\thanks{I thank Alessandro Bacchetta and Asmita Mukharjee for many insightful discussions during the DIS'17 workshop.}\\
        Department of Physics, Indian Institute of Technology Kanpur, Kanpur 208016, India\\
        E-mail: \email{tanmay@iitk.ac.in}}

\author{Dipankar Chakrabarti\\
Department of Physics, 
Indian Institute of Technology Kanpur,
Kanpur 208016, India\\
        E-mail: \email{dipankar@iitk.ac.in}}

\abstract{We present model predictions of Collins asymmetries for proton in the SIDIS process for $\pi^+$ and $\pi^-$ channels. We use a recently proposed light-front quark-diquark model for proton with the light-front wave functions predicted in soft-wall AdS/QCD. The model predictions are compared with the HERMES and COMPASS data. }

\FullConference{XXV International Workshop on Deep-Inelastic Scattering and Related Subjects\\
		3-7 April 2017\\
		University of Birmingham, UK}

\begin{document}
\section{Introduction\label{intro}}
In the past few decades, several experimental collaborations have measured Collins single spin asymmetry. This asymmetry provides correlation between parton transverse polarization in a
transversely polarized nucleon and transverse momentum of the produced hadron which indicate existence of non-vanishing transverse momentum of interior quarks and collinear picture is no longer sufficient to describe the transverse structure of nucleons. 

Many phenomenologycal models have addressed this interesting phenomenon \cite{Boffi09}. Most of the model calculations consider Gaussian Anstz for TMDs and FFs to extract the corresponding distribution functions from the fitting of the asymmetry data. A simultaneous extraction of Collins function and transversity distribution is proposed by Anselmino $et. al.$ \cite{Ans09,Ans13} from Collins asymmetry  data of HEREMS and COMPASS.  

The QCD factorization scheme for Semi-Inclusive Deep Inelastic Scattering(SIDIS) allow to factorize the cross-section for the one photon exchange process $\ell N \to \ell' h X$ as 
\be 
d\sigma^{\ell N \to \ell' h X}=\sum_\nu \hat{f}_{\nu/P}(x,\bfp;Q^2)\otimes d\hat{\sigma}^{\ell q \to \ell q} \otimes \hat{D}_{h/\nu}(z,\bfk;Q^2);
\ee  
where $\hat{f}_{\nu/P}(x,\bfp;Q^2)$  represents  the transverse momentum dependent parton distribution functions(TMDs),
$d\hat{\sigma}^{\ell q \to \ell q}$ represents the heard scattering which is a point like QED scattering mediated by a virtual photon and the third term and $\hat{D}_{h/\nu}(z,\bfk;Q^2)$ is for fragmentation functions(FFs).
Such a scheme holds in small $\bfPhp$ and large $Q$ region, $P_{h\perp}^2 \simeq \Lambda^2_{QCD} \ll Q^2 $. At large $\bfPhp$, quark-gluon corrections and higher order pQCD corrections become important.
In the $\gamma-N$ center of mass frame, the kinematical variables are defined  as 
$ x=\frac{Q^2}{2(P.q)}=x_B, ~~
y=\frac{P.q}{P.\ell}=\frac{Q^2}{s x}$ and $
z=\frac{P.P_h}{P.q}=z_h. $
We use the light-cone coordinates $x^\pm = x^0 \pm x^3 $ and  momentum of the incoming proton is $P \equiv (P^+, \frac{M^2}{P^+}, \textbf{0}_\perp)$. The struck quark of momentum $p \equiv (xP^+, \frac{p^2+|\bfp|^2}{xP^+}, \bfp)$ interact with a virtual photon and transferred the energy of amount $Q^2=-q^2$. The fractional energy transferred by the photon in the lab system is $y$ and the energy fraction carried by the produced hadron is $z=P_h^-/k^-$. The momentum carried by the produced hadron is $\bfPh \equiv (P^+_h, P^-_h, \bfPhp) $.
In this frame, though the incoming proton dose not have transverse momentum, the constituent quarks can have  non-zero transverse momenta which sum up to zero. $\bfp, \bfk$ and $\bfPhp$ are the transverse momentum carried by struck quark, fragmenting quark and fragmented hadron respectively. The relation between them, at $\mathcal{O}(\bfp/Q)$, is given by 
$
\bfk=\bfPhp-z\bfp.
$
The polarised SIDIS cross-section is written in terms of structure functions as \cite{Ans11}
\be 
\frac{d\sigma^{\ell(S_\ell)+P(S_P)\to \ell' P_h X}}{dx_B dy dz d^2\bfPhp d\phi_S} &=& \frac{2\alpha^2}{s x y^2}\bigg\{\frac{1+(1-y)^2}{2}\bigg\}F_{UU}+...\nonumber\\
&& + S_T\bigg[(1-y)(\sin(\phi_h+\phi_S)F^{\sin(\phi_h+\phi_S)}_{UT}+...\bigg]+...
\ee 
 at  $\mathcal{O}(\bfp/Q)$. Where the proton polarization vector is defined as $S_P \equiv (S_T \cos \phi_S, S_T \sin \phi_S, S_L)$ and the quark polarization vector is defined as $ S^q \equiv (S^q_T \cos \phi_{S^q}, S^q_T \sin \phi_{S^q}, S^q_L)$. The weighted structure functions, $F^{w(\phi_h,\phi_S)}_{S_\ell S}$ is written as a convolution of TMDs and FFs as
\be 
F^{w(\phi_h,\phi_S)}_{S_\ell S}&=&\mathcal{C}[w \hat{f}(x,\bfp) \hat{D}(z,\bfk)] \nonumber\\
&=& \sum_\nu e^2_\nu  \int d^2\bfp d^2\bfk \delta^{(2)}(\bfPhp-z\bfp-\bfk) w(\bfp,\bfPhp) \hat{f}^\nu(x,\bfp)\hat{D}^\nu(z,\bfk),\label{conv}
\ee
where  $\hat{f}^\nu(x,\bfp)$ and $\hat{D}^\nu(z,\bfk)$ represent leading twist TMDs and FFs respectively. 
The Collins asymmetry is written as
\be 
A^{\sin(\phi_h+\phi_S)}_{UT}(x,z,\bfPhp,y)&=& 2\frac{\int d\phi_h d\phi_S [d\sigma^{\ell P^\uparrow \to \ell' h X}-d\sigma^{\ell P^\downarrow\to \ell' h X}]\sin(\phi_h+\phi_S)}{\int d\phi_h d\phi_S [d\sigma^{\ell P^\uparrow \to \ell' h X}+d\sigma^{\ell P^\downarrow\to \ell' h X}]} \nonumber \\
&=& \frac{4\pi^2\alpha^2\frac{(1-y)}{s x y^2} F^{\sin(\phi_h+\phi_S)}_{UT}(x,z,\bfPhp,y)  }{2\pi^2\alpha^2\frac{1+(1-y)^2}{s x y^2}F_{UU}(x,z,\bfPhp,y)} \nonumber\\
&=& \frac{4\pi^2\alpha^2\frac{(1-y)}{s x y^2} \mathcal{C}\bigg[\frac{P_{h\perp}-z (\hat{\bf{P}}_{h\perp}.\bfp)}{zM_h} h^{\nu}_{1}(x,\bfp^2) H^{\perp\nu}_1(z,\bfk^2)\bigg] }{2\pi^2\alpha^2\frac{1+(1-y)^2}{s x y^2}\mathcal{C}[ f^\nu_1(x,\bfp^2) D^{h/\nu}_1(z,\bfk^2)]}.\label{Coll_Asy}
\ee
Where the $P^\uparrow$ and $P^\downarrow$ represents the transverse polarization of proton along positive and negative directions respectively. 

\section{Model Calculations }\label{sec_SSA_model}
We predict the Collins asymmetry by calculating the unpolarised TMDs and transversity TMDs in a recently proposed light-front quark diquark model(LFQDM) \cite{LFQDM} for the SIDIS process $\ell N \to \ell^\prime X h$ at $\mu^2=2.5~GeV^2$ and compare with the experimental data measured by COMPASS and HERMES for $\pi^+$ and $\pi^-$ channels. 
In this model, we assume that one of the three valence quarks of nucleon interacts with the virtual photon and other two valence quarks form a diquark of spin-0(Scalar Diquark) or spin-1(Vector diquarks). The proton state is written as two particle bound state of a quark and a diquark having a spin-flavour $SU(4)$ structure.
 \be 
|P; \pm\rangle = C_S|u~ S^0\rangle^\pm + C_V|u~ A^0\rangle^\pm + C_{VV}|d~ A^1\rangle^\pm. \label{PS_state}
\ee 
Where $\mid u~ S^0\rangle,~|u~ A^0\rangle$ and $|u~ A^0\rangle$ are two particle states having isoscalar-scalar, isoscalar-axial vector and isovector-axial vector diquark respectvely \cite{Jakob97_Bacc08}. The states are written in two particle Fock state expansion with $J^z =\pm1/2$ for both the scalar and the axial vector diquarks \cite{LFQDM}. The two particle fock state wave functions are adopted from soft-wall AdS/QCD prediction \cite{BT} and modified as
\be
\varphi_i^{(\nu)}(x,\bfp)=\frac{4\pi}{\kappa}\sqrt{\frac{\log(1/x)}{1-x}}x^{a_i^\nu}(1-x)^{b_i^\nu}\exp\bigg[-\delta^\nu\frac{\bfp^2}{2\kappa^2}\frac{\log(1/x)}{(1-x)^2}\bigg].
\label{LFWF_phi}
\ee
 We use the AdS/QCD scale parameter $\kappa =0.4$ GeV as determined in \cite{CM1} and the quarks are  assumed  to be  massless. The parameters $a_i^\nu,b_i^\nu$ and $\delta^\nu$ are fixed by fitting the Dirac and Pauli form factors. 

In the light-front formalism, the TMDs correlator at equal light-front time $z^+=0$  is defined for SIDIS as
\be
\Phi^{\nu [\Gamma]}(x,\textbf{p}_{\perp};S)&=&\frac{1}{2}\int \frac{dz^- d^2z_T}{2(2\pi)^3} e^{ip.z} \langle P; S|\overline{\psi}^\nu (0)\Gamma \mathcal{W}_{[0,z]} \psi^\nu (z) |P;S\rangle\Bigg|_{z^+=0} \label{TMD_cor}
\ee
 for different Dirac structures $\Gamma=\gamma^+,\gamma^+\gamma^5$ and $i\sigma^{j+}\gamma^5$. 
We choose a frame where the proton is collinear with photon and the final hadron has transverse momentum.
The proton spin components are $S^+ = \lambda_N \frac{P^+}{M},~ S^- = \lambda_N\frac{P^-}{M},$ and $ S_T $ with helicity $\lambda_N$.

In the LFQDM,  the explicit form of  $f_1$ and $h_1$ TMDs are given by
\be 
{f}^{\nu  }_1(x,\bfp)&=&\bigg(C^2_SN^{\nu 2}_S+C^2_V\big(\frac{1}{3}N^{\nu 2}_0+\frac{2}{3}N^{\nu 2}_1\big)\bigg)\frac{\ln(1/x)}{\pi\kappa^2}\bigg[T^\nu_1(x) +\frac{\bfp}{M^2} T^\nu_2(x)\bigg]\exp\big[-R^\nu(x)\bfp^2\big],\label{TMD_f1}\\
{h}^{\nu  }_1(x,\bfp) &=& \bigg(C^2_SN^{\nu 2}_S-C^2_V\frac{1}{3}N^{\nu 2}_0\bigg) \frac{\ln(1/x)}{\pi\kappa^2}T^\nu_1(x)\exp\big[-R^\nu(x)\bfp^2\big], \label{TMD_h1}
\ee
\be
{\rm where},~ T^\nu_1(x)= x^{2a^{\nu}_1}(1-x)^{2b^{\nu}_1-1}; \hspace{0.3cm}
T^\nu_2(x)= x^{2a^{\nu}_2 -2}(1-x)^{2b^{\nu}_2-1}; \hspace{0.3cm}
R^\nu(x)=\delta^\nu \frac{\ln(1/x)}{\kappa^2 (1-x)^2}.\label{Fx}
\ee
The values of the model parameters $a^\nu_i, b^\nu_i(i=1,2)$ and $\delta^\nu$ are given in \cite{LFQDM} at initial scale $\mu_0=0.8~GeV$.
The pre-factors containing $C_j$ and $N_k(j,k=S,0,1)$ are the normalized constants which satisfy the quark counting rules for unpolarised TMDs.

We use Gaussian anstz for fragmentations functions as discussed in Ref. \cite{Ans09,Ans13}. 
 The values of the parameters are listed in \cite{Ans13} and  $D^{h/\nu}_1(z)$ is taken from the phenomenological extraction \cite{Kret01}. 

\section{Collins asymmetry in LFQDM }
In this model, using the TMDs from Eqs.(\ref{TMD_f1}-\ref{TMD_h1}) and FFs from \cite{Ans13,Kret01} 
the explicit expression for Collins asymmetry reads as  
\be 
A^{\sin(\phi_h+\phi_S)}_{UT}&=&  \frac{\frac{2(1-y)}{s x y^2} \frac{P_{h\perp}\sqrt{2 e}}{M_h} \sum_\nu e^2_\nu  \hat{h}^{\nu}_1(x)\mathcal{N}^C_\nu(z) D^{h/\nu}_1(z)\frac{\avksq ^2_C \avpsq_x}{\avksq\avPhsq_C}\frac{e^{-\bfPhp^2/\avPhsq _C}}{\avPhsq_C}}{\frac{1+(1-y)^2}{s x y^2}\sum_\nu e^2_\nu N^\nu_{f_1}\frac{\ln(1/x)}{\pi\kappa^2}\bigg[T^\nu_1(x)+
 \frac{\langle m^2_\perp \rangle}{M^2} T^\nu_2(x)\bigg]D^{h/\nu}_1(z)\avpsq_x \frac{e^{-\bfPhp^2/\avPhsq}}{\avPhsq}}, \label{Coll_LFQDM}
\ee
\be 
{\rm where},~\hat{h}^{\nu}_1(x)=& \bigg(C^2_SN^{\nu 2}_S-C^2_V\frac{1}{3}N^{\nu 2}_0\bigg) \frac{\ln(1/x)}{\pi\kappa^2}T^\nu_1(x),  & \avpsq_x = 1/R(x)= \frac{\kappa^2 (1-x)^2}{\delta \log(1/x)},\\
 N^\nu_{f_1}=& \bigg( C^2_SN^{\nu 2}_S+C^2_V\big(\frac{1}{3}N^{\nu 2}_0 + \frac{2}{3}N^{\nu 2}_1\big)\bigg),  & \avksq_C = \frac{M_h^2 \avksq}{M_h^2 + \avksq},\\
\langle m^2_\perp \rangle =& \bigg[\avksq \avPhsq + z^2 P_{h\perp}^2 \avpsq_x\bigg]\frac{\avpsq_x}{\avPhsq^2}, &\avPhsq_C  = \avksq_C + z^2 \avpsq_x.
\ee
\begin{figure}
\begin{center}
\includegraphics[scale=0.5]{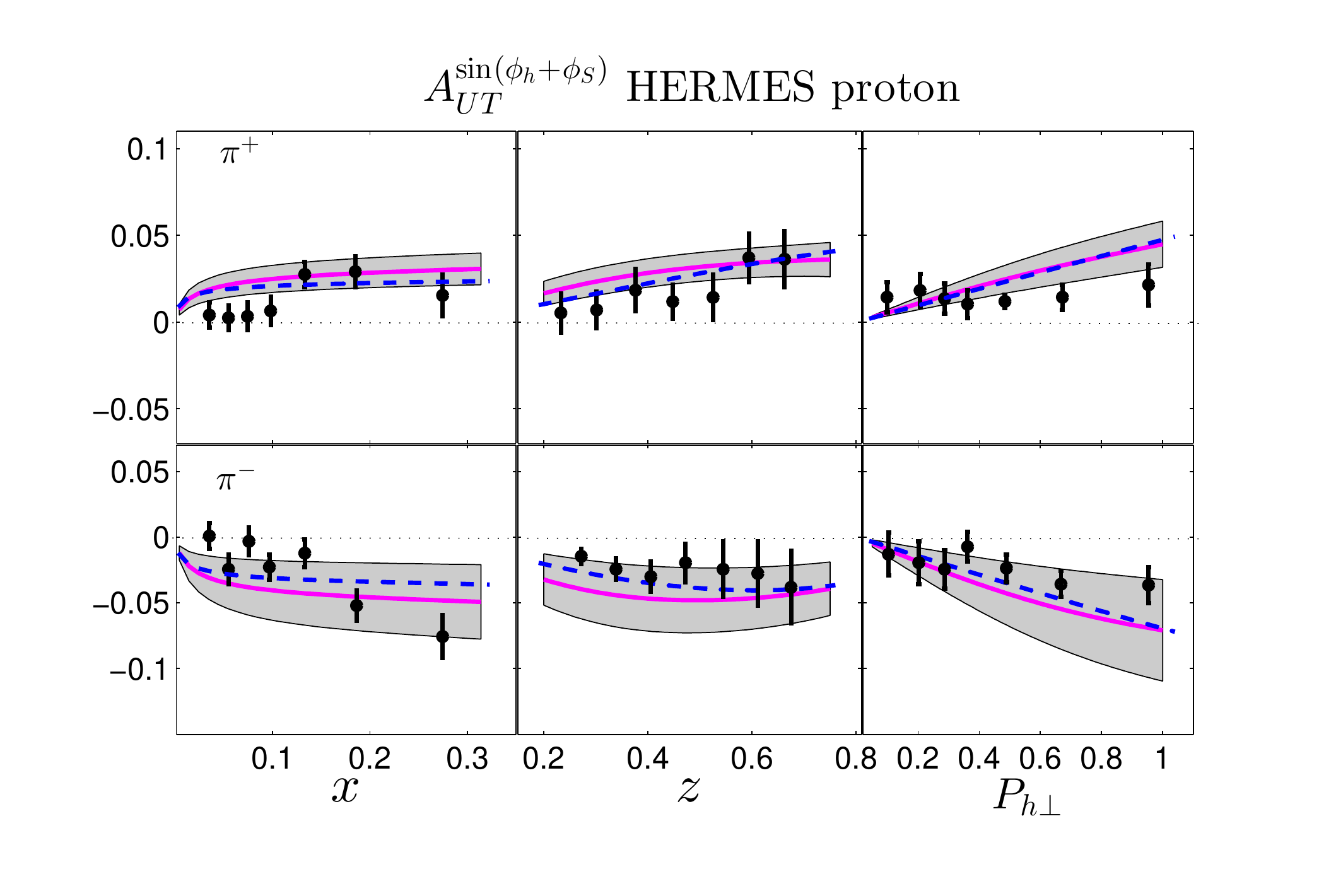} 
\end{center}
\caption{\label{fig_Col_H} (Color online) Model prediction to Collins asymmetry within HERMES kinematics are presented and compared with experimental data by HERMES Collaboration \cite{HERMES10} for $\pi^+$(upper row) and $\pi^-$(lower row) channels. The continuous(pink) lines with error regions(gray) represent the results corresponding to the QCD evolution of TMDs \cite{Aybat11,Ans12} and the dashed(blue) lines represent the results where the TMDs are evolved in parameter evolution approach \cite{LFQDM}.}
\end{figure}
\begin{figure}
\begin{center}
\includegraphics[scale=0.5]{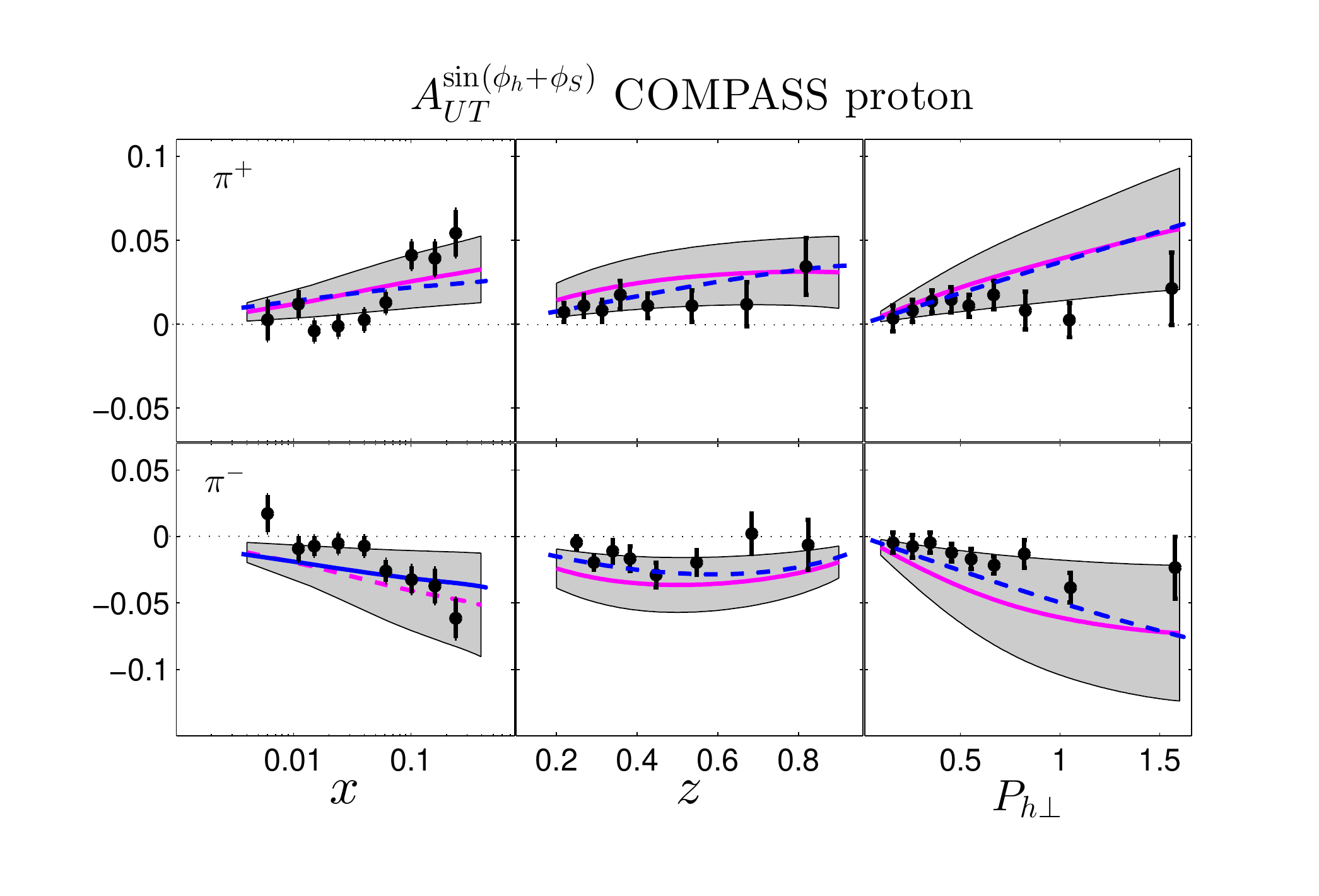} 
\end{center}
\caption{\label{fig_Col_C}(Color online) Model prediction to Collins asymmetry within COMPASS kinamatics are presented and compared with experimental data by COMPASS Collaboration \cite{COMPASS14} for $\pi^+$(upper row) and $\pi^-$(lower row) channels. The symbols and linestyle are same as used in Fig.\ref{fig_Col_H}.} 
\end{figure}
Since the COMPASS and HERMES experiments measure the the asymmetries at higher scale, scale evolution of the TMDs and FFs are needed to have a model prediction to the experimental data. 
Here we use the LO parametrization for fragmentation functions at the scale $2.5 ~GeV^2$( which is the average $Q^2$ values of the HERMES experiment). 
Therefore, for consistency, we evolve the TMDs to the scale $\mu^2=2.5 ~GeV^2$ to give a model prediction for Collins asymmetry. We perform the evolution of the TMDs(Eq.(\ref{TMD_f1}-\ref{TMD_h1})) in two different approaches: one is the reduced QCD evolution approach discussed in  \cite{Ans12} and other one is the parameter evolution approach of LFQDM proposed in \cite{LFQDM, TMD_LFQDM}. In this model, we compare the QCD evolutions generated with two different kernels: the kernel $\tilde{R}(\mu,\mu_0,b_T)$ and the reduced kernel $R(\mu,\mu_0)$ which is defined as $R(\mu,\mu_0) \equiv \tilde{R}(\mu,\mu_0,b_T\to \infty)$ as discussed in \cite{Ans12} and found a very insignificant difference between them. So, we use the reduced kernel to solve the evolution equation analytically.

 The Collins asymmetry written in Eq.(\ref{Coll_LFQDM}) is a function of the variables $x, z, y$ and $\bfPhp$ at a given scale $\mu$. Since experimentally the asymmetries are measured with one variable at a time, one has to integrate the denominator and numerator of Eq.(\ref{Coll_LFQDM}) separately over all the other variables within the corresponding kinematical limits.
Our model predictions to the Collins asymmetries are shown in Fig.{\ref{fig_Col_H}, for the HERMES kinematics $0.023 \leq x \leq 0.4,~~ 0.2 \leq z \leq 0.7 $ and $0.1 \leq y \leq 0.95$, at the scale $\mu^2=2.5 ~ GeV^2$ and compare with HERMES data for $\pi^+$ and $\pi^-$ channels. 
 The first, second and third columns are the variation with $x,z$ and $P_{h\perp}$ respectively. The upper row is for $\pi^+$ channel and the lower row is for $\pi^-$ channel. The continuous lines(pink) represent our model results when the TMDs are evolved in reduced QCD evolution and the dashed lines(blue) are the model result when TMDs ate evolved in parameter evolution approach. The gray errorbars represent the error corresponding to the continuous(pink) lines. A large contribution to the errors comes from the uncertainties of the parameters of FFs \cite{Ans13}. Error corridor corresponding to the parameter evolution approach is not included to avoid clumsiness in the figure. 
Note that the evolution of the parameters in LFQDM are consistent with the DGLAP evolution of unpolarised PDFs. Therefore the generalization of this evolution approach  to TMDs may not give satisfactory results for individual TMD evolution. But  for the case of Collins asymmetry which involves the ration of the TMDs,   the common contributions in the numerator and denominator get partially cancelled and give quite satisfactory results. 

 In Fig.{\ref{fig_Col_C}, we also present our model results for Collins asymmetries within the COMPASS kinematics  $0.003 \leq x \leq 0.7,~~ 0.2 \leq z \leq 1.0 $ and $0.1 \leq y \leq 0.9$, at the scale $\mu^2=2.5 ~ GeV^2$  and compare with the experimental findings by the COMPASS collaboration.
 

%

\section{Conclusions}
Here we have presented the results of Collins asymmetries for proton in a SIDIS process using a light-front quark-diquark model where the wave functions are constructed in the AdS/QCD prediction. Our results have reasonably good agreements with the HERMES and COMPASS data within the error bars except for the higher values of $P_{h\perp}$. 
The evolution of Collins asymmetry is provided by the evolution of TMDs and we present the model prediction for reduced QCD evolution approach and the parameter evolution approach. 
The proposal of the parameter evolution approach shows reasonably good results like the QCD evolution approach in both the kinematical regime. We found a positive asymmetry for $\pi^+$ channel and negative asymmetry for $\pi^-$ channel as observed in experimental measurements. It will be interesting to investigate other azimuthal spin asymmetries in this model and compare with experimental data.



\begin{thebibliography}{99}
\bibitem{Boffi09} 
  S.~Boffi, A.~V.~Efremov, B.~Pasquini and P.~Schweitzer,
  Phys.\ Rev.\ D {\bf 79}, 094012 (2009);
  M.~Anselmino, M.~Boglione, U.~D'Alesio, A.~Kotzinian, F.~Murgia, A.~Prokudin and C.~Turk,
  Phys.\ Rev.\ D {\bf 75}, 054032 (2007);
  A.~Bacchetta, L.~P.~Gamberg, G.~R.~Goldstein and A.~Mukherjee,
  Phys.\ Lett.\ B {\bf 659}, 234 (2008)

\bibitem{Ans09} 
  M.~Anselmino, M.~Boglione, U.~D'Alesio, A.~Kotzinian, F.~Murgia, A.~Prokudin and S.~Melis,
  Nucl.\ Phys.\ Proc.\ Suppl.\  {\bf 191}, 98 (2009)
\bibitem{Ans13} 
  M.~Anselmino, M.~Boglione, U.~D'Alesio, S.~Melis, F.~Murgia and A.~Prokudin,
  Phys.\ Rev.\ D {\bf 87}, 094019 (2013)
 
\bibitem{Ans11} 
  M.~Anselmino, M.~Boglione, U.~D'Alesio, S.~Melis, F.~Murgia, E.~R.~Nocera and A.~Prokudin,
  Phys.\ Rev.\ D {\bf 83}, 114019 (2011)
  

\bibitem{LFQDM} 
  T.~Maji and D.~Chakrabarti,
  Phys.\ Rev.\ D {\bf 94}, no. 9, 094020 (2016)

\bibitem{Jakob97_Bacc08}
  R.~Jakob, P.~J.~Mulders and J.~Rodrigues,
  Nucl.\ Phys.\ A {\bf 626} (1997) 937;
  A.~Bacchetta, F.~Conti and M.~Radici,
  Phys.\ Rev.\ D {\bf 78} (2008) 074010.

  
\bibitem{BT}
  S.~J.~Brodsky and G.~F.~de Teramond,
  Phys.\ Rev.\ D {\bf 77} (2008) 056007;
  G.~F.~de Teramond and S.~J.~Brodsky,
  arXiv:1203.4025 [hep-ph]. 

\bibitem{CM1}D.~Chakrabarti and C.~Mondal,
  Phys.\ Rev.\ D {\bf 88}, no. 7, 073006 (2013);
  Eur.\ Phys.\ J.\ C {\bf 73}, 2671 (2013).
  
\bibitem{Kret01}
  S.~Kretzer, E.~Leader and E.~Christova,
  Eur.\ Phys.\ J.\ C {\bf 22}, 269 (2001)

\bibitem{Ans12} 
  M.~Anselmino, M.~Boglione and S.~Melis,
  Phys.\ Rev.\ D {\bf 86}, 014028 (2012). 
  
\bibitem{TMD_LFQDM} 
   T.~Maji and D.~Chakrabarti,
  Phys.\ Rev.\ D {\bf 95}, no. 7, 074009 (2017)

 \bibitem{HERMES10} 
  A.~Airapetian {\it et al.} [HERMES Collaboration],
  Phys.\ Lett.\ B {\bf 693}, 11 (2010)

\bibitem{Aybat11} 
  S.~M.~Aybat and T.~C.~Rogers,
  Phys.\ Rev.\ D {\bf 83}, 114042 (2011).
  

  
\bibitem{COMPASS14} 
  A.~Martin [COMPASS Collaboration],
  Phys.\ Part.\ Nucl.\  {\bf 45}, 141 (2014)


\end{thebibliography}
\end{document}